\documentclass[aps,prl,twocolumn,superscriptaddress]{revtex4-1}

\usepackage[pagewise]{lineno}

\usepackage{amsmath}
\usepackage{graphicx}   
\usepackage{breqn}
\usepackage{natbib}
\usepackage{amssymb}

\begin{document}

\newcommand{\simtel}{\textit{sim\_telarray}}
\newcommand{\impact}{{ImPACT}}

\title{Systematic Differences due to High Energy Hadronic Interaction Models in Air Shower Simulations in the 100\,GeV-100\,TeV Range}

\author{R.D. Parsons}
\thanks{daniel.parsons@mpi-hd.mpg.de}

\author{H. Schoorlemmer}
\affiliation{Max-Planck-Institut f\"ur Kernphysik P.O. Box 103980$,$ D 69029 Heidelberg$,$ Germany}

\begin{abstract}

The predictions of hadronic interaction models for cosmic-ray induced air showers contain inherent uncertainties due to limitations of available accelerator data and theoretical understanding in the required energy and rapidity regime. Differences between models are typically evaluated in the range appropriate for cosmic-ray air shower arrays ($10^{15}$-\,$10^{20}$\,eV). However, accurate modelling of charged cosmic-ray measurements with ground based gamma-ray observatories is becoming more and more important.

We assess the model predictions on the gross behaviour of measurable air shower parameters in the energy (0.1-100\,TeV) and altitude ranges most appropriate for detection by ground-based gamma-ray observatories. We go on to investigate the particle distributions just after the first interaction point, to examine how differences in the micro-physics of the models may compound into differences in the gross air shower behaviour. Differences between the models above 1\,TeV are typically less than 10\%. However,  we find the largest variation in particle densities at ground at the lowest energy tested (100\,GeV), resulting from striking differences in the early stages of shower development.

\end{abstract}

\maketitle

\section{Introduction}

\begin{figure*}[]
\begin{center}
\includegraphics[width=0.99\textwidth]{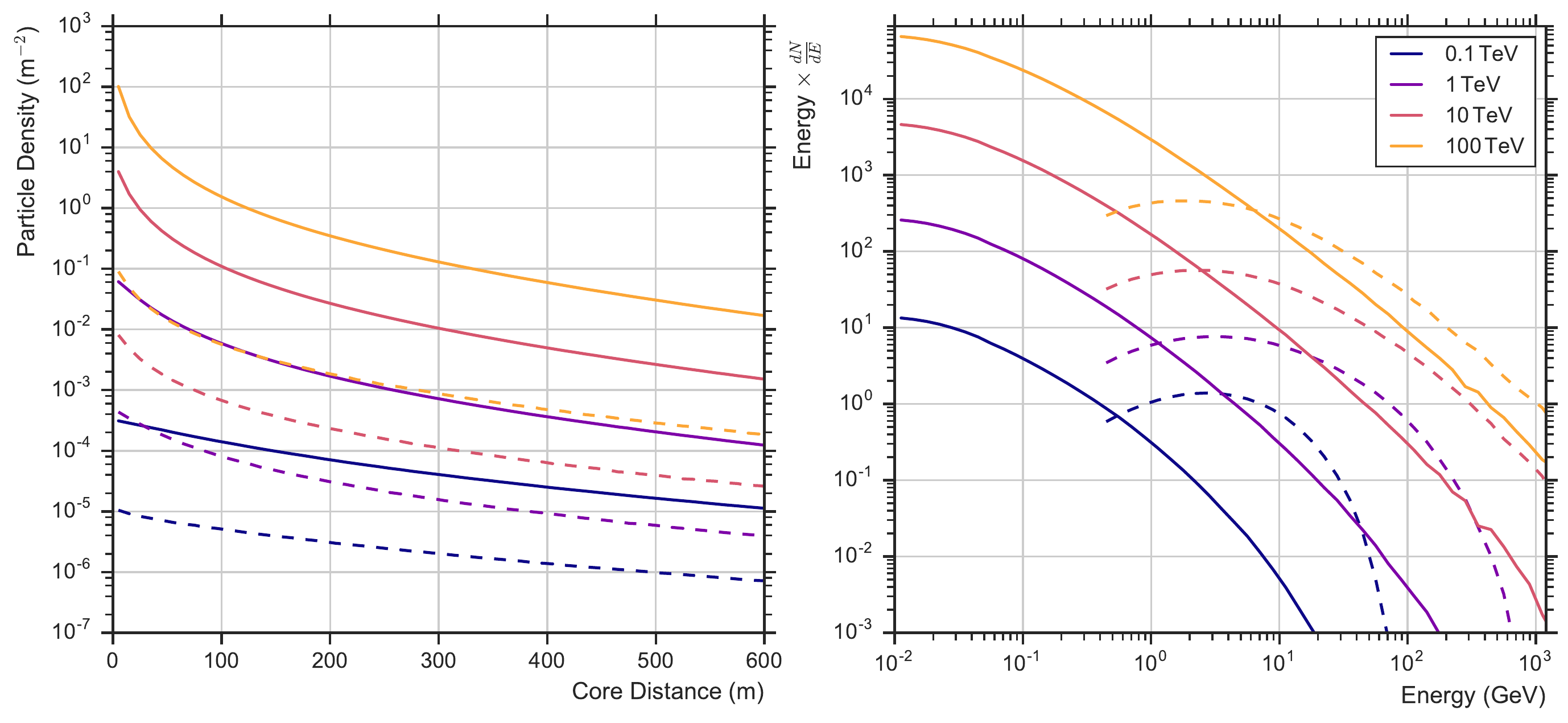}
\caption{Lateral distribution functions in number density for the EM component (solid)  and muon component(dashed) [left] and spectral density [right] at ground (4100\,m) for vertical simulated showers with EPOS-LHC hadronic interaction model. Cuts are placed on the particle energy of 0.3 MeV on electrons and gammas and 300 MeV on muons. }
\label{fig-partLDF}
\end{center}
\end{figure*}

Ground based gamma-ray astronomy uses the particle shower initiated
when a very high energy ($>100$\,GeV) gamma-ray interacts with an
atom in the Earth's atmosphere, to detect and reconstruct the
primary particles of the primary gamma-rays. Such reconstruction is
performed either by directly detecting the energetic particles formed
in the shower at ground level (as in the HAWC gamma-ray observatory \cite{HAWC_CRAB})
or detecting the Cherenkov light emitted as the
relativistic particles pass through the atmosphere using Imaging Atmospheric Cherekov Telescopes (IACTs), as in the H.E.S.S.,
MAGIC and VERITAS experiments \cite{HESS_Performance,MAGIC_Performance, VERITAS_Performance}.
Event reconstruction in all these instruments is performed in combination
with detailed Monte Carlo simulations of the air shower development,
which is typically very well understood for gamma-ray induced air
showers. 

Both experiment types also observe the flux from air showers 
induced from hadronic cosmic rays, which constitutes the major
background for gamma-ray ray observations. 
Although they can be largely rejected, some cosmic-ray contamination
will remain which must be estimated. This can be problematic as lack
of knowledge of microscopic hadronic physics in the energy range of
interest, leads to significant variations in the air shower
predictions. In most observations of gamma-ray sources this uncertainty can be negated
by instead using positions within the instrument field of view which
contain no gamma-ray emission to estimate the background
contamination (e.g. \cite{berge}). However, in the case of extremely large or truly
diffuse sources, no such signal free region exists and the background 
contamination may need to be estimated from simulated data (like for example 
the Fermi-Bubbles~\cite{FermiBubble} and the Halo around 
Geminga~\cite{HAWCGeminga} in the case of IACTs).

This is even more apparent in non gamma-ray observations, as most pointedly
seen in the measurement of the cosmic ray electron spectrum 
\cite{HESS_electrons, HESS_electrons2, VERITAS_electrons, MAGIC_electrons} where the background
contamination must be estimated from comparison to Monte Carlo 
simulations, in which case the systematic uncertainties in the 
hadronic interactions quickly become the dominant form of uncertainty 
in the measurement. Or in the case when these observatories are used to perform 
measurement of the hadronic cosmic rays (for example \cite{HAWCCRSpec}).
Finally, hadron simulations are required to estimate the sensitivity of future
gamma-ray observatories, such as the Cherenkov Telescope Array (CTA) \cite{CTA}, 
the Southern Gamma-ray Survey Observatory (SGSO) \cite{SGSO}, and the Large High Altitude Air Shower Observatory (LHAASO) \cite{LHAASO} which could be affected by the choice of models.

The advent of LHC measurements has provided large amounts of data at
never before probed energies \cite{LHCmod} and at extreme rapidities
\cite{LHCF1, LHCF2}. This data glut promises improvements to hadronic
interaction models by facilitating model tuning in energy ranges not
before possible and has resulted in the creation of a new generation
of air shower focused hadronic interaction models
\cite{EPOS-LHC,QGSJetII-04,Sibyll2.3c} tuned to this data set.

\begin{figure*}[]
\begin{center}
\includegraphics[width=0.97\textwidth]{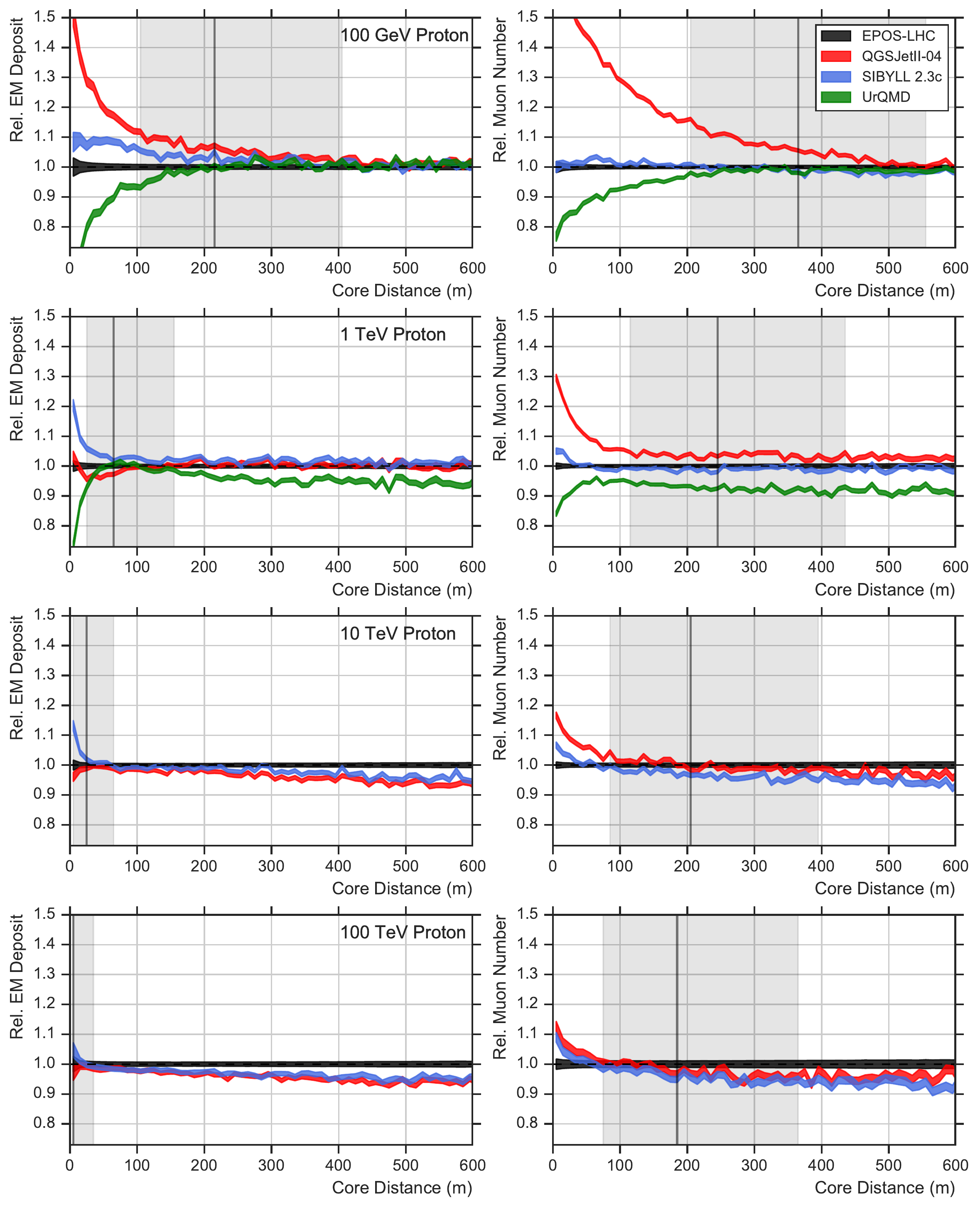}
\caption{Lateral Distribution Functions for the four particle energies tested (increasing top to bottom) at 4100\,m altitude relative to  predictions from EPOS-LHC (width of the line represents the error on the mean). Panels on the left show relative energy deposit per unit area of EM particles, while panels on the right show relative number of muons per unit area. The energy of the primary proton increases from top to bottom panels. The grey shaded area shows the region of 50\% containment centred on the median (vertical line).}
\label{fig-partLDFFrac}
\end{center}
\end{figure*}

Although many detailed tests have been performed on this model
generation, 
most have concentrated on the predictions in the energy range from
$10^{15}$\,eV to $10^{19}$\,eV \cite{PAO}, where measurements such as the primary
particle mass can be critically dependent on the model used. Such
studies lie far beyond the energy range of interest for gamma-ray
instruments from $10^{11}$\,eV to $10^{14}$\,eV.  In order to investigate the model behaviour in this
energy regime, we have therefore studied the effects of the most
commonly used hadronic interaction model versions included in the air shower
simulation package CORSIKA \cite{corsika}, concentrating on model
predictions relevant to ground-based gamma-ray astronomy with both
atmospheric Cherenkov telescopes and particle detectors at observation 
levels appropriate to these detector types. 

In this publication we concentrate on understanding the gross 
behaviour of the shower, in terms of the particles or Cherenkov light arriving
at ground level. We compare the spread of
predictions in this energy range, quantifying the level of systematic
uncertainties in measurements due to the use of hadronic interaction
models and testing the hypothesis that the tuning of models to a wider
range of data will inevitably lead to a convergence of model predictions. 
We then go on to compare the particle production of the first interaction to understand how the differences in air shower 
development stem from differences in the interaction characteristics of individual particles.
It should be noted that the average distributions presented here do not necessarily 
correspond directly to measurables at gamma-ray observatories and biases in 
observations might expose different systematic differences between the model predictions.

\section{Air Shower Simulations}

\begin{table}[]
\begin{tabular}{|c||c|c|c|}
\hline
\textbf{Energy}  & \textbf{Particle} & \textbf{Cherenkov} & \textbf{First} \\ 
  & \textbf{Distributions} & \textbf{Distributions} & \textbf{Interaction} \\ 
\hline
\hline

{100\,GeV} & 10$^6$ & 10$^6$ & 10$^6$ \\
{1\,TeV}   & 10$^5$ & 10$^5$ & 10$^6$ \\
{10\,TeV}  & 10$^4$ & 10$^4$ & 10$^6$ \\
{100\,TeV} & 10$^3$ & 10$^3$ & 10$^6$ \\ \hline
\end{tabular}
\label{tab-eventStats}

\caption{Number of events simulated at different energies for the studies presented.}
\end{table}

To test the prediction of the models in gamma-ray observatories
a series of simulation sets were created using CORSIKA with the
Cherenkov light option activated. Protons showers were simulated at
zenith, with CORSIKA v7.64 using (at the time of writing) the latest
versions of the high energy hadronic interaction models EPOS-LHC, QGSJet-II-04 and
SIBYLL 2.3c (assumed to be valid at energies $>$80\,GeV, the default transition energy in CORSIKA).
Each of the above mentioned high energy hadronic interaction 
models was combined with the UrQMD \cite{UrQMD} low energy hadronic 
interaction model with the cross over energy set at 80\,GeV. The
electromagnetic shower component was simulated with the EGS4 \cite{EGS4}
model. 
Energy cuts were placed on both electrons and muons of 0.3 and 300\,MeV respectively within the CORSIKA simulations.
Finally to better understand the transition between high end low
energy interaction models, showers were also simulated using only the UrQMD 
model in the lowest energy bands tested.
To best calculate the relevant parameters for the 
two types of gamma-ray observatories currently operating the ground level of the simulations were set to 4100\,m (which corresponds the elevation of the HAWC gamma-ray observatory) and 1800\,m (corresponding to the elevation of the  H.E.S.S. experiment). Protons make up the dominant source of background in these experiments, hence no simulations were performed with heavier nuclei.
Vertical showers were simulated in the typical 
energy range of the background of gamma-ray observations at 4 fixed
energies (100\,GeV, 1\,TeV, 10\,TeV and 100\,TeV) for both observation levels, 
the number of simulated events is detailed in table~\ref{tab-eventStats}.

\subsection{Ground Level Particle Distributions}
\label{sec:latdist}

\begin{figure*}[]
\begin{center}
\includegraphics[width=0.99\textwidth]{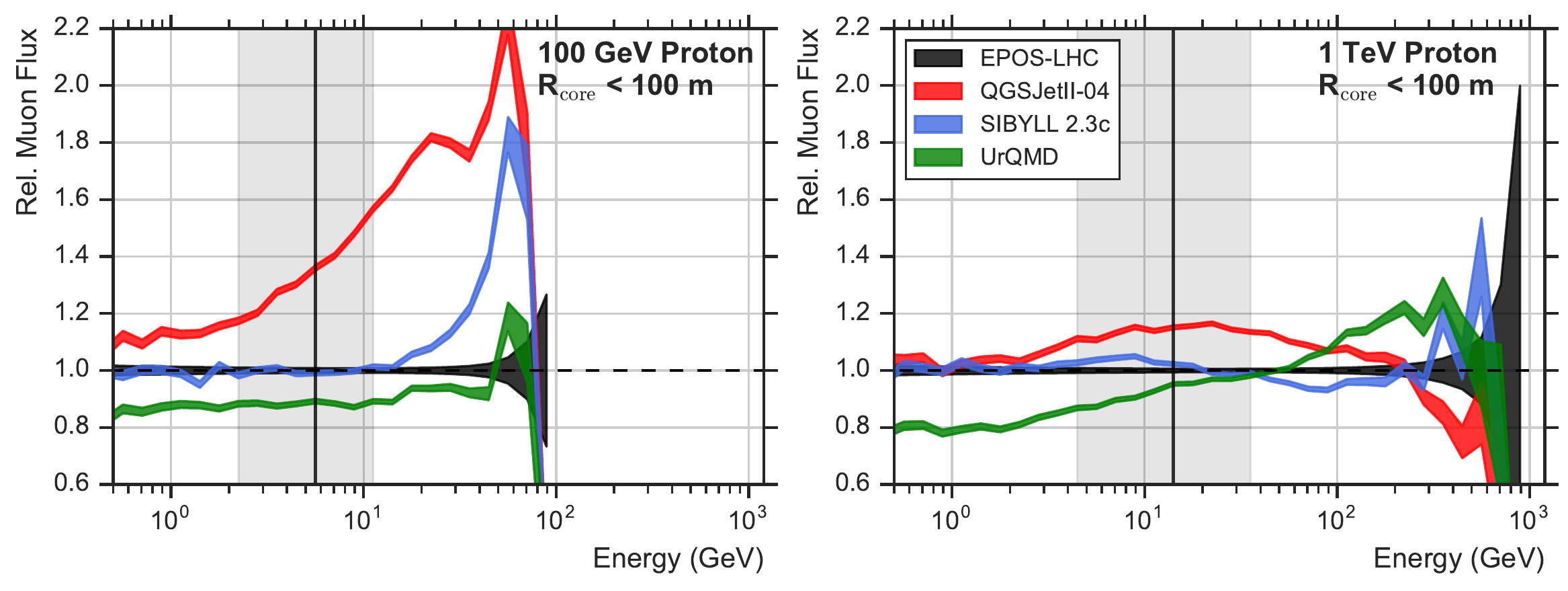}
\caption{Spectrum of muons falling within 100\,m from the shower core at 4100\,m altitude shown relative to EPOS-LHC 
(width of the line represents the error on the mean). 
The grey shaded area shows the region of 50\% containment centred on the median (vertical line).}
\label{fig-muonSpecFrac}
\end{center}
\end{figure*}

Figure \ref{fig-partLDF} shows the averaged lateral distribution function (LDF) 
of particles (left) and the particle spectrum (right)
at 4100\,m altitude for the EPOS model. This figure demonstrates the both the 
comparative steepness of the LDF of electrons and gammas when compared to the muon LDF
with the electromagnetic component being more concentrated toward the shower core. The
spectra show that most muons lie in the range from 1 to 10\,GeV, while most electrons 
and gammas lie at somewhat lower energies.


Figure \ref{fig-partLDFFrac} shows the fractional deviation of QGSJet, SIBYLL and UrQMD (for 100\,GeV and 1 TeV only) from EPOS for both the sum of energy deposited at ground level by photons, electrons and positrons (hereafter EM energy deposit) [left]
and the muon number [right].  The choice of parameters roughly mimics the behaviour of a HAWC-like obsevatory, where the EM component of the air shower generates a particle cascade within the detector, depositing all of its energy, whereas muons simply pass through depositing a fixed amount of energy. 
The trend in the relative values of both energy deposit and muon number is clearly seen to be evolving as a function of energy, with the deviations being largest 
at the lowest energies. At all energies, the difference between the models in both the 
energy deposit and muon number seems to be largest
in the region near the shower axis. 

At 100\,GeV SIBYLL shows a 10\% larger energy deposit than EPOS at 10\,m from the shower core.
When using QGSJet this difference becomes even larger with more than 40\% increase in
energy deposit and a more than 50\% larger muon density below 10\,m.
Although some extreme differences are seen in the LDF shape, the differences
in total particle production remains rather small, with QGSJet producing only 10\%
larger total EM deposit and 10\% more muons. UrQMD on the other hand shows 
the opposite behaviour as QGSJet, showing a 30\% reduction in EM energy deposit
and 20\% lower muon number close to the shower core.

Such large deviations in LDF shape for the different models
at 100\,GeV is rather surprising as the
energy range of interactions within these showers lies well within the
range of energies described by existing accelerator data. This result is made
all the more surprising when one considers the transition energy
between the high and low energy hadronic models at 80\,GeV (all
simulations share UrQMD as low energy model), so it seems quite likely
that the differences in the high energy models arise dominantly from the differing treatment of the
first interaction(s) of the shower.

At 1\,TeV and 10\,TeV the differences are somewhat smaller, showing a
spread in the predicted energy deposit of 5\% or less and a similar spread 
of the  muon density. However, at small impact distances the significant 
increase in relative particle number at ground of 
QGSJet remains, most clearly seen in the muon number. Similarly the
lower EM energy deposit and muon number seen in UrQMD is still seen at a reduced level at 1\,TeV.
The 10 and 100\,TeV simulations show a similar consistency between models, 
however interestingly the order of
the relative production level of muons has reversed, with QGSJet and
SIBYLL now producing around a few percent smaller EM deposit and muons number than EPOS at all impact distances over 50\,m.

\begin{figure*}[t]
\begin{center}
\includegraphics[width=0.99\textwidth]{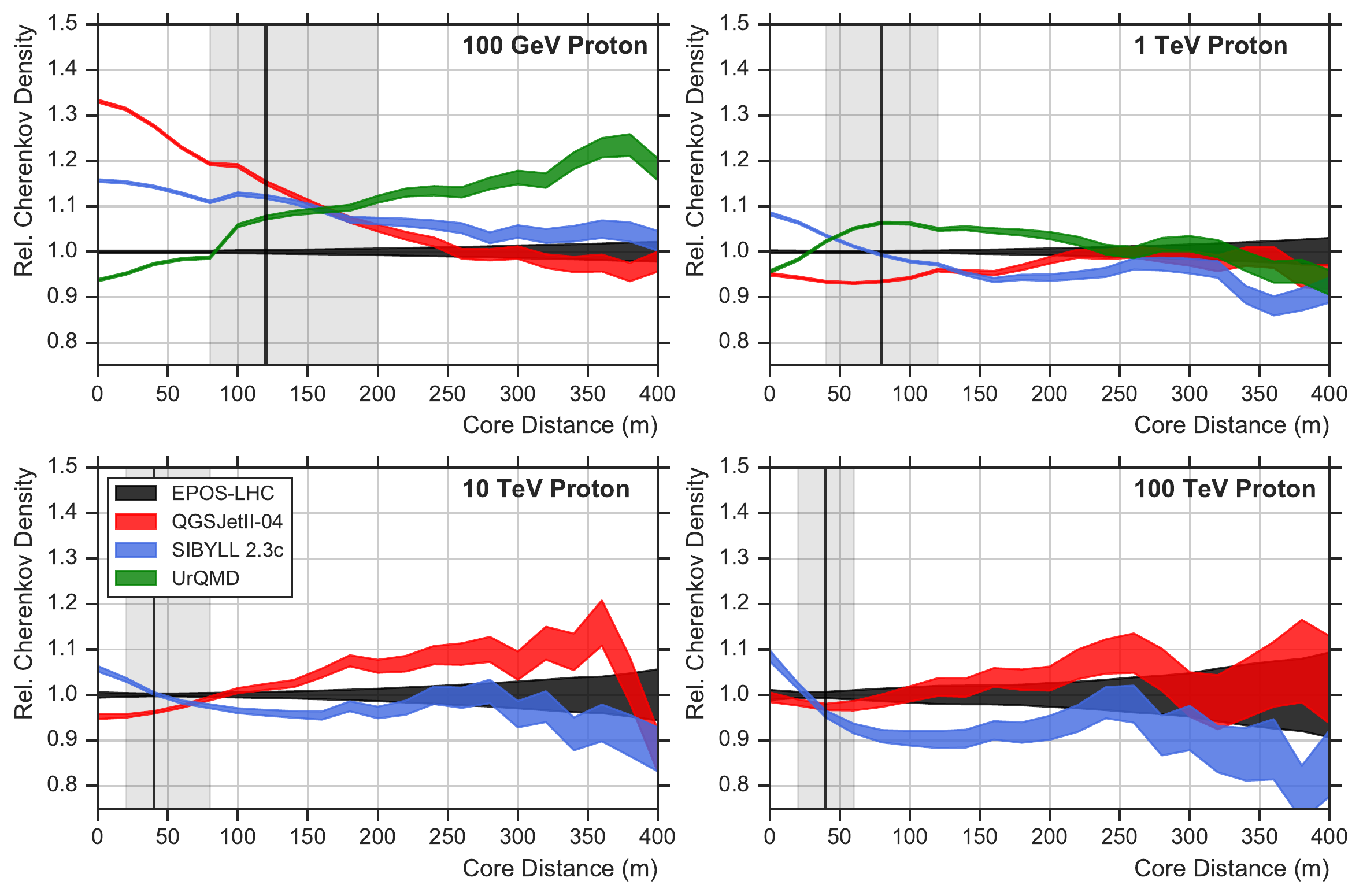}
\caption{Relative number density of Cherenkov 
photons in the range 300-600\,nm after the application of atmospheric absorption 
at 1800\,m altitude relative to EPOS-LHC (width of the line represents the error on the mean). 
The grey shaded area shows the region of 50\% 
containment centred on the median (vertical line). }
\label{fig-CherenkovLDF}
\end{center}
\end{figure*}

When one considers the spectrum of muons at ground level in the central 
100\,m from the shower cores (figure \ref{fig-muonSpecFrac}) 
it becomes clear that the excess of muons seen close to the shower core at 100\,GeV
is caused by an excess of energetic ($>$2\,GeV) muons.
These energetic muons
are certainly produced in the earliest interactions in the shower
development and are therefore quite indicative of a significant difference in the 
pion production spectrum in the 100\,GeV range (investigated in later sections). 
UrQMD shows a similar increase in the number of energetic muons, however when one considers
the energy range where most muons lie, UrQMD  shows a deficit of around 10\%.
As primary energy increases the difference 
in the high energy muon spectra decreases. Already at 1\,TeV the model discrepancies are significantly reduced, most clearly in the highest energy muons. For example at 1\,TeV QGSJet now produces around 10\% more muons in the peak energy range.

\subsection{Cherenkov Photon Distribution}


To evaluate the impact of hadronic interaction models on the intensity of Cherenkov light seen by IACTs we also simulate the lateral
distribution of Cherenkov photons arriving at an observation level of
1800\,m with the CORSIKA standard atmospheric absorption tables applied. 
The relative normalisation of the Cherenkov LDF defines the energy scale of the detected cosmic ray induced air showers.
In contrast to air shower array detection the Cherenkov light originates all altitudes within the air shower (although atmospheric absorption reduces the contributions from high altitudes), therefore relates to the full air shower development rather than relying on only the energetic particles that make it to ground-level. 

Figure \ref{fig-CherenkovLDF} shows the comparison of lateral photon
distributions between the interaction models from 100\,GeV to 100\,TeV. 

At 100\,GeV, like in the ground-level particle distributions (figure \ref{fig-partLDFFrac}), the largest spread in model prediction is seen.
In addition, the ordering of the relative Cherenkov light below 75\,m impact distance is the same as the EM energy deposit. While for the EM energy deposit all the models seem to converge at large impact distances, this is not the case for the Cherenkov density, where UrQMD shows the largest deviation at larger core distance. 

At 1, 10 and 100\,TeV the relative behaviour of the models again change. 
The LDF looks similar in all models, differences between the models fluctuate around the 5\% level. 

\section{Early Shower Development}

\begin{figure*}[]
\begin{center}
\includegraphics[width=0.99\textwidth]{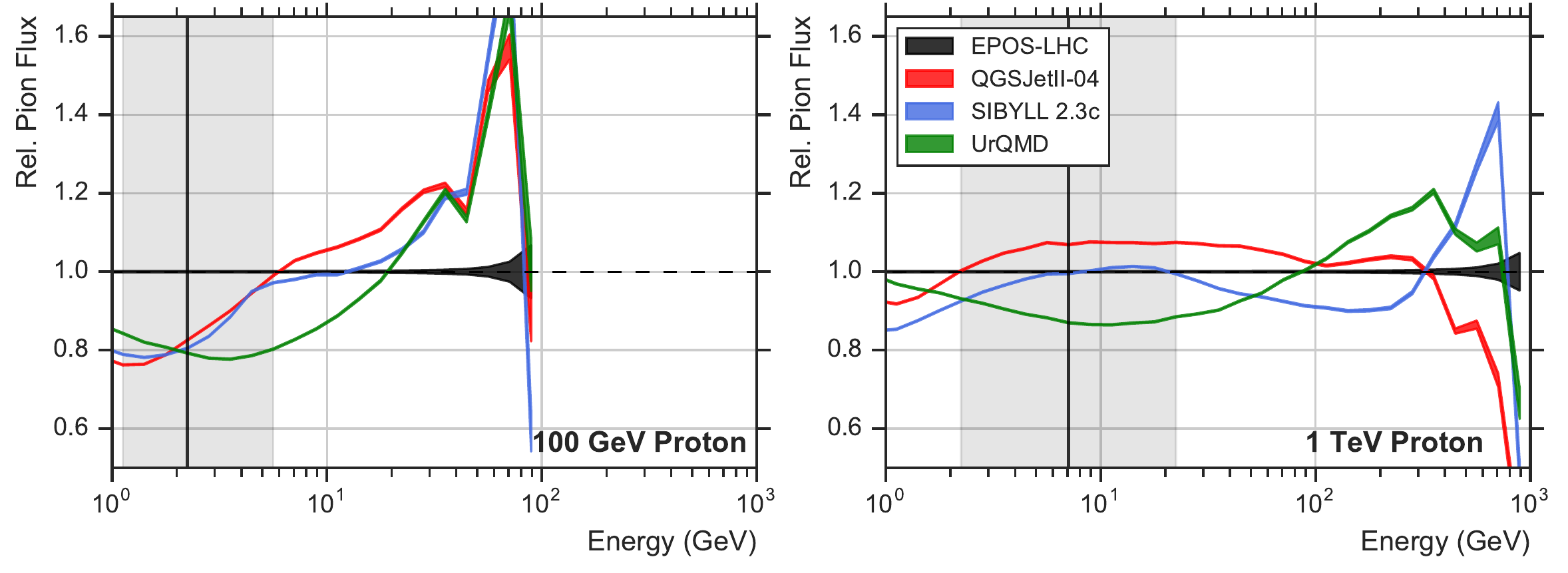}
\caption{Comparison of the charged pion energy spectrum  (shown as a function of energy) relative to EPOS-LHC from the first proton Nitrogen interaction in the air shower (width of the line represents the error on the mean).}
\label{fig-pion-spec}
\end{center}
\end{figure*}

The differences in the air shower prediction are most obvious at 100\,GeV primary energy. However, the energy where the simulation switches between low and high energy interaction models at 80\,GeV. Therefore differences must already occur in the very early stages of shower development.   
To investigate the first interaction more thoroughly, simulations were generated a fixed interaction of primary proton with a Nitrogen nucleus. We evaluate the particle distributions 1\,cm below the interaction point.

As the deviations are most apparent in the muon distributions \ref{fig-partLDFFrac}, we evaluate the distributions of charged pions as their decay is the dominant channel for muon production. Figure \ref{fig-pion-spec} shows the relative comparison of the pion energy spectrum for different hadronic interaction models. 

Significant differences between the model predictions are seen in the pion production at all energies but are again most apparent at 100\,GeV. For pions around 1\,GeV EPOS produces around 20\% more muons than the other high and low energy interaction models, while in the range 50-100\,GeV between 30 and 60\% more muons are seen.

As illustrated by the relative pion energy spectrum for 1\,TeV protons, at higher primary particle energies differences between the models are typically below 15\%, with the exception when the energy transfer to charge pions approaches the energy of the primary proton.

\begin{figure*}[]
\begin{center}
\includegraphics[width=0.99\textwidth]{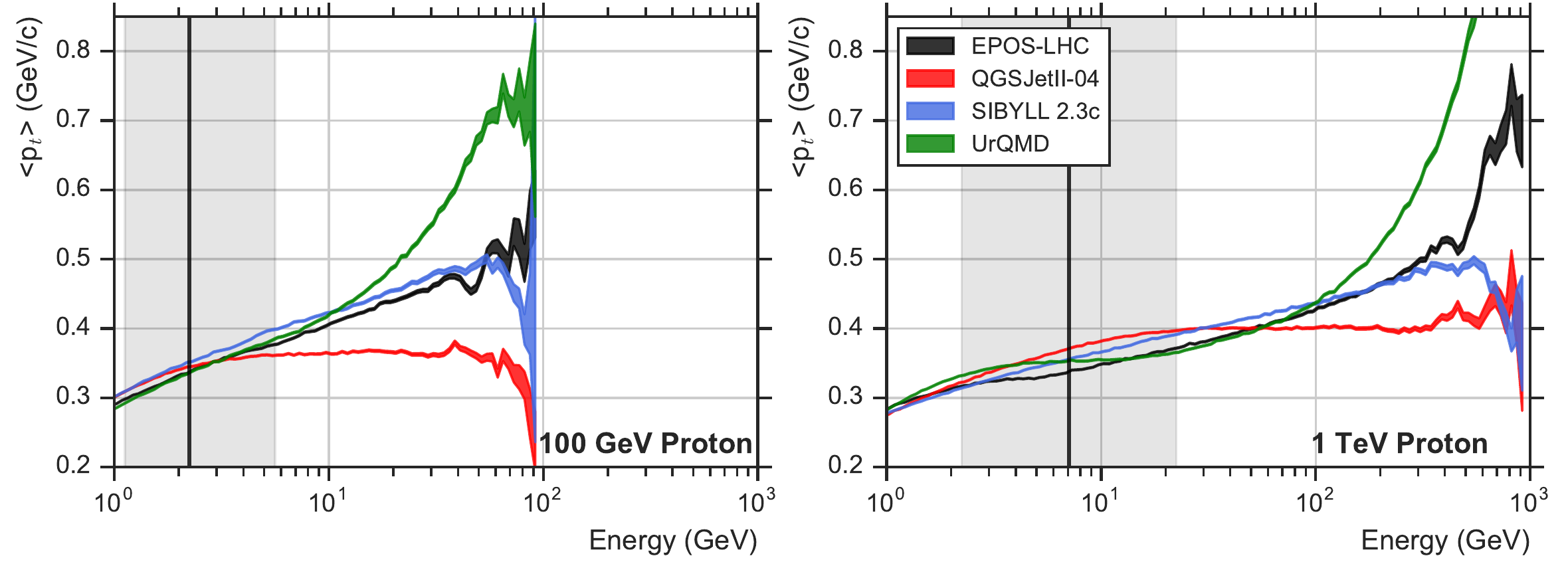}
\caption{Comparison of the mean transverse momentum transfer of charged pions (shown as a function of energy) from the first proton Nitrogen interaction in the air shower (width of the line represents the error on the mean).}
\label{fig-pt}
\end{center}
\end{figure*}

Finally, figure \ref{fig-pt} shows the average transverse momentum to secondary pions
as a function of the pion energy. Across the proton energy range
at  energies less than 5\% of the primary energy there is an excellent agreement between all models. However,
at all incident proton energies the transverse momentum of secondary pions above 10\% energy fraction 
seems to diverge. Again this is most apparent at 100\,GeV primary energy with EPOS and SIBYLL producing roughly consistent predictions which increase with energy fraction. QGSJet on the other hand seems to flatten out, imparting around 25\% less transverse momentum, while UrQMD imparts around 25\% more to the most energetic pions. At 1,TeV the reduced p$_{\rm t}$ transfer is still present in QGSJet, but EPOS now seems to peak in p$_{\rm t}$ transfer for the most energetic particles. Similar behaviour is observed for the gammas produced shortly after the first interaction.

\section{Discussion}

The major differences between the air shower predictions are summarised by the following points:

\begin{itemize}

\item The largest differences in air shower prediction are seen at 100\,GeV, decreasing with energy

\item At low energies QGSJet shows a significantly steeper lateral distribution of both the EM energy deposit and muon number than SIBYLL and EPOS, while UrQMD shows the opposite behaviour and has significant flatter lateral distributions. 

\item The excess muons close to the shower core in QGSJet are mostly energetic particles

\item The lateral distribution of Cherenkov photons shows a steeper behaviour in QGSJet and SIBYLL at low energies than EPOS, the opposite behaviour is seen in UrQMD

\end{itemize}

The differences in the behaviour of the 100\,GeV and 1\,TeV showers can be largely understood by the dominance of the behaviour of the first interaction at this energy. Showers in this energy range are not well defined with only a few generations of particles being produced since the majority of the pions produced in the hadronic interactions (those that only carry a small fraction of the particle energy) quickly decay without further interaction. As the first interactions occur high in the atmosphere, the decay products from these pions are unable to reach ground level, also Cherenkov light from such particles are strongly absorbed.

Therefore, it is clear that when measuring low energy air showers at ground level the most important events are those where a large fraction of the primary particle energy is passed to a single pion. Such events have an important impact on the ground level particle distributions for two main cases. 
In the case that such a pion undergoes another hadronic interaction, the start of shower development is effectively postponed, pushing further particle production closer to ground level. Whereas energetic charged pions that do not interact further will produce energetic muons that are able to reach the ground without decaying (for a 20\,km pion decay altitude typically muons below around 3\,GeV will decay before reaching ground level).

In this context the air shower predictions can be better understood by considering the behaviour of the first interactions. One can see in figures~\ref{fig-pion-spec}~\&~\ref{fig-pt} that the behaviour of low energy ($<$5\% proton energy) pions is quite consistent between the models. However, for the more important energetic pions the results are quite different. At both 100\,GeV and 1\,TeV QGSJet produces more (charged) pions in the 5-50\,GeV range than SYBILL and EPOS while UrQMD produces fewer. Such an increased number of energetic pions in QGSJet could potentially be explained by the extreme defferences in the $\rho^0$ production spectrum shown in \cite{SPSrho0}.
This difference in pion production number likely correlated the relative enhancement and deficit in muons numbers seen in QGSJet and UrQMD respectively. In addition, the regime in the transverse momentum distribution where are single pion gets a large fraction of the primary energy, helps to concentrate particles in the shower core for QGSJet whilst widening the particle distribution for UrQMD.

As discussed above,  at the lowest energies the shower observables are biased towards showers that develop deeply in the atmosphere, however as the energy increases this is no longer the case. Therefore the influence of the first interaction in shower development becomes less and less important with increasing primary energy for the ground level observables. 
In addition, as the primary particle energy increases the differences in the pion production spectrum at the first interaction point between the models generally become smaller.
However the differences in the production spectrum at low energies can still play an important role in governing the particle distribution at ground as these low energy interactions now represent the most numerous interactions in the shower and now take place closer to the detector level.

\section{Conclusions}

In this paper we have demonstrated that, a good agreement (10\% level) is seen between air shower predictions from different hadronic interaction models at energies above 10\,TeV. However, contrary to typical assumption, the predictions in gross air shower behaviour do not converge at lower energies but rather diverge in the simple predictions tested. The primary reason for this divergence in behaviour seems to be chiefly related to the early stage of shower development and the differing predictions in energy and momentum distributions of pions. 

It is clear from this comparison that even though these showers lie at the lowest boundary of the model validity range, more tuning to accelerator data is required to reproduce the average air shower behaviour at this low energy limit. However, it seems quite likely that accelerator data may not be readily available in the relevant psuedorapidity range.
In order to better tune these models we should also attempt to leverage the air shower data from current and future gamma-ray detectors to provide further cosmic ray measurements for the comparison of model results. 
Although the gross average air shower behaviour studied here is not strictly representative of the most important measurables used in gamma-ray instruments in this energy range, it is indicative for the overall shower behaviour. Additionally the large differences seen between model behaviours suggest it will be possible to construct observables with gamma-ray observatories which provide useful input for model tuning and will help to improve the reliability of predictions in future.
Such tests might be developed for air shower arrays such as HAWC or future facilities such as SGSO, however some assumptions of the cosmic ray composition would have to be made. Muon LDF and production height measurements should also be possible with the Cherenkov telescope array \cite{MitchellMuon}, however measurements in the 100\.GeV to 1\,TeV range may be challenging. Particularly the combination of ground particles and Cherenkov light measured by LHAASO may help to better distinguish between model predictions.


\section{Acknowledgements}

The authors thank the MPIK non-thermal astrophysics group for fruitful discussions about the paper, in particular J.A. Hinton for providing helpful comments. We would also like to thank T. Pierog and S. Ostapchenko for useful suggestions for the manuscript.


\bibliographystyle{elsarticle-num_etal} 
 \bibliography{./references}

\end{document}